\documentstyle[12pt]{article}
\begin{document}
\newcommand{\newc}{\newcommand}

\newc{\be}{\begin{equation}}
\newc{\ee}{\end{equation}}
\newc{\ba}{\begin{eqnarray}}
\newc{\ea}{\end{eqnarray}}
\newc{\ie}{{\it i.e.}}
\newc{\eg}{{\it eg.}}
\newc{\etc}{{\it etc.}}
\newc{\etal}{{\it et al.}}

\newc{\ra}{\rightarrow}
\newc{\lra}{\leftrightarrow}
\newc{\no}{Nielsen-Olesen }
\newc{\lsim}{\buildrel{<}\over{\sim}}
\newc{\gsim}{\buildrel{>}\over{\sim}}

\begin{titlepage}
\begin{center}
May 1993\hfill
             \hfill
\vskip 1in

{\large \bf
On the Asymptotics of Nielsen-Olesen Vortices
}

\vskip .4in
{\large Leandros Perivolaropoulos}\footnote{E-mail address:
leandros@cfata3.harvard.edu},\footnote{Also, Visiting Scientist,
Department of Physics, Brown University, Providence, R.I. 02912.}\\[.15in]

{\em Division of Theoretical Astrophysics\\
Harvard-Smithsonian Center for Astrophysics\\
60 Garden St.\\
Cambridge, Mass. 02138, USA.}
\end{center}
\vskip .2in
\begin{abstract}
\noindent
We investigate analytically and numerically the asymptotic behavior of the
Nielsen-Olesen vortex solutions and show that they approach their asymptotic
values exponentially but with exponents that differ from the ones quoted in
the literature. In particular, it is shown that for values of the Higgs
self-coupling that are larger than a critical value, both the Higgs field
and the gauge field approach exponentially their asymptotic values with
{\it equal}
exponents that are {\it independent} of the Higgs self-coupling.
 \end{abstract}

\end{titlepage}

\par
The classic paper of Nielsen-Olesen \cite{no73} on the existence of vortex
solutions in the Abelian-Higgs model has generated significant activity on
both particle physics and cosmology during the past two decades. One
of the main results of that paper was the proof of the existence of
vortex solutions in the Abelian-Higgs model and the derivation of their
asymptotic behavior as a function of the parameters in the Lagrangian.
 This asymptotic behavior has been taken for granted in subsequent papers
\cite{vort}
and quoted to be the same as that obtained in the \no paper. However,
as we show here, for values of the Higgs self coupling
larger than a critical value the asymptotic behavior of the vortex solutions
is not the one quoted by \no.

 The Abelian-Higgs model Lagrangian may be written as:
\be
L=-{1\over 4} F^{\mu \nu}F_{\mu \nu} +{1\over 2} \vert D_\mu \Phi \vert ^2
-{\lambda \over 4} (\vert \Phi \vert ^2 - \eta ^2)^2
\ee
where $\Phi$ is a complex scalar (the Higgs field),
$D_\mu = \partial_\mu - ie A_\mu$ and $F_{\mu \nu}=\partial_\mu A_\nu-
\partial_\nu A_\nu$.

Using the \no anzatz:
\ba
\Phi&=&\eta f(r) e^{i\theta}\\
A_\mu&=&{\hat e}_\theta {v(r)\over {er}}
\ea
and rescaling the coordinate r, we obtain the rescaled \no
equations for the dimensinless functions $f(r)$ and $v(r)$:
\ba
f''(r)+{1\over r} f'(r)-{{(1-v(r))^2}\over r^2} f(r) -
\beta (f(r)^2 - 1) f(r)&=&0\\
v''(r)-{1\over r} v'(r) + 2 (1-v(r))f(r)^2 &=&0
\ea
where $\beta = {\lambda \over {e^2}}=({{\mu_H} \over {\mu_A}})^2$
where $\mu_H$ and $\mu_A$ are the masses of the Higgs and
the gauge field respectively. The boundary conditions are
\ba
f(r)&\ra & 0 \\
v(r)&\ra & 0
\ea
for $r\ra 0$ and
\ba
f(r)&\ra & 1\\
v(r)&\ra & 1
\ea
for $r\ra \infty$.

The rate by which $f(r)$ and $v(r)$ are approaching their asymptotic
value at $r\ra \infty$ may be obtained by using the ansatz:
\ba
f&\ra & 1+\delta f \\
v&\ra & 1+\delta v
\ea
and keeping only lowest order terms in $\delta f$ while keeping all terms in
$\delta v$. The resulting equations are:
\ba
\delta f'' +{1\over r} \delta f' - {{(\delta v)^2}\over r^2}
-2 \beta \hspace{.2cm} \delta f &=& 0\\
\delta v'' - {1 \over r} \delta v' - 2 \delta v &=& 0
\ea

Using an ansatz of the form
\be
\delta v = e^{-\gamma r} r^\alpha (c^v _1 + r^{-1} c^v _2)
\ee
in (13) and equating the coeficients of $e^{-\gamma r} r^\alpha$ and
$e^{-\gamma r} r^{\alpha -1}$ to 0, we obtain
\be
\delta v \ra c^v _1 e^{-\sqrt{2} r} r^{1/2}
\ee
where $c^v_1, c^v_2$ are constants of $O(1)$.
Substituting now (15) into (12) and
using an ansatz similar to (14)  for $\delta f$ we
may obtain both the particular and the homogeneous solution
of (12) valid for $r\gg 1$. The result is:
\be
\delta f \ra \delta f_h + \delta f_p \equiv
c^f {{e^{-\sqrt{2\beta} r}}\over \sqrt{r}} - {{(c^v_1)^2 e^{-2\sqrt{2}r}}
\over {2(\beta - 4) r}}
\ee
or
\ba
\delta f &\ra & c^f {{e^{-\sqrt{2\beta} r}}\over \sqrt{r}} \hspace{3cm}
\beta \lsim 4 \\
\delta f &\ra &  - {{(c^v_1)^2 e^{-2\sqrt{2}r}}
\over {2(\beta -4) r}} \hspace{3cm} \beta>4
\ea
where $c^f$ is a constant of $O(1)$. The case $\beta < 4$ is in agreement with
the asymptotic behavior obtained by \no and subsequent papers up to a factor of
${1\over \sqrt{r}}$.  This factor is usually not quoted in the literature (with
the exception of Ref. \cite{v79}) even though it can be of some significance.

For $\beta=4$, (16) implies that $c_1^v \rightarrow 0$ for finiteness, which
justifies the `$\beta \lsim 4$' in (17).

The case $\beta > 4$ indicates a significant
modification that should be imposed
in the results of \no. In this case, the exponent in (18) is fixed and is
independent of $\beta$ while according to \no and later papers
the exponent should
be $\sqrt{2\beta}$ (same as for $\beta < 4$). The reason that this point was
missed is that the term ${{(\delta v)^2}\over r^2}$,
being second order in $\delta v$ was ignored for all values of $\beta$. Clearly
however, this term can not be ignored for $\beta >4$. The condition $\beta > 4$
physiaclly means that the Higgs mass $\mu_H$ is beyond the two vector meson
threshold ($\mu_H > 2\mu_A$)

We have checked the validity of (17), (18) by numerically solving the system
(4), (5)
using collocation at gaussian points (a variation of the relaxation
scheme). Fig. 1 shows a plot of $ln({\delta f(r)})\over r$ vs $r$ for
\be
\beta=({1\over 2}, 2, 8, 50)
\ee
 Clearly, $ln({\delta f(r)})\over r$ is
approaching an asymptotic value for all $\beta$ which is in agreement
with the exponents in (17), (18). In fact, the asymptotic values corresponding
to (19) according to Fig. 1 are
\be
(\gsim -1.1,\gsim -2.05,\gsim -3.1, \gsim -3.3)
\ee
In comparison, Refs \cite{no73}, \cite{vort} would predict the asymptotic
values
$(-1,-2,-4,-10)$ while our analysis predicts $(-1,-2,-2.82,-2.82)$
(since $2\sqrt{2}\simeq 2.82$) which is in much better agreement with the
numerical result shown in (20).

In conclusion, we have shown that the width of the scalar part
of the \no vortex is independent of the Higgs self-coupling $\beta$ for
$\beta > 4$ and that for these values of $\beta$
it is equal to the width of the
gauge part of the vortex.  This result, which was missed in previous studies,
may have interesting consequences in applications of the \no vortices. Such
applications include the stability
of embedded strings \cite{stab}, the existence of vortex solutions
in two-Higgs systems, the interactions of vortices \etc. Investigation of
these effects is currently in progress.

\bigskip
\newpage

{\bf Acknowledgements}

\noindent
I am grateful to T. Vachaspati for
interesting discussions and to S. Garcia for his help in the
numerics.
This work was supported by a CfA Postdoctoral Fellowship.
\vskip 1cm
\centerline{\large \bf Figure Captions}

{\bf Figure 1:}
The dependence of $ln({\delta f(r)})\over r$ on $r$ for $\beta=({1\over 2},
2,8,50)$. The asymptotic values should be compared with the
predicted exponents in (17), (18).

\end{document}